\documentclass[12pt]{iopart}

\usepackage{graphicx}
\usepackage{subfigure}  
\begin{document}

\title[]{Searching for the QCD Critical Point Using Particle Ratio Fluctuations and Higher Moments of Multiplicity Distributions}

\author{Terence J Tarnowsky (for the STAR Collaboration)}

\address{National Superconducting Cyclotron Laboratory,
Michigan State University, East Lansing, MI 48824, USA}
\ead{tarnowsk@nscl.msu.edu}
\begin{abstract}
Dynamical fluctuations in global conserved quantities such as baryon number, strangeness, or charge may be observed near a QCD critical point. Results from new measurements of dynamical $K/\pi$, $p/\pi$, and $K/p$ ratio fluctuations are presented. The commencing of a QCD critical point search at RHIC has extended the reach of possible measurements of dynamical $K/\pi$, $p/\pi$, and $K/p$ ratio fluctuations from Au+Au collisions to lower energies. The STAR experiment has performed a comprehensive study of the energy dependence of these dynamical fluctuations in Au+Au collisions at the energies $\sqrt{s_{NN}}$ = 7.7, 11.5, 39, 62.4, and 200 GeV. New results are compared to previous measurements and to theoretical predictions from several models. The measured dynamical $K/\pi$ fluctuations are found to be independent of collision energy, while dynamical $p/\pi$ and $K/p$ fluctuations have a negative value that increases toward zero at top RHIC energy. Fluctuations of the higher moments of conserved quantities (net-proton and net-charge) distributions, which are predicted to be sensitive to the presence of a critical point, are also presented.

\end{abstract}


\section{Introduction}

Fluctuations and correlations are well known signatures of phase transitions. In particular, the quark/gluon to hadronic phase transition may lead to significant fluctuations \cite{Koch1}. In 2010, the Relativistic Heavy Ion Collider (RHIC) began a program to search for the QCD critical point. This involves an ``energy scan'' of Au+Au collisions from top collision energy ($\sqrt{s_{NN}}$ = 200 GeV) down to energies as low as $\sqrt{s_{NN}}$ = 7.7 GeV \cite{STARBES}. This critical point search will make use of the study of correlations and fluctuations, particularly those that could be enhanced during a phase transition that passes close to the critical point \cite{Stephanov, kurtosis}. 

$\nu_{dyn}$ quantifies deviations in the particle ratios from those expected for an ideal statistical Poissonian distribution \cite{nudyn1, nudyn2}. The definition of $\nu_{dyn,K/\pi}$ (describing fluctuations in the $K/\pi$ ratio) is,
\begin{eqnarray}
\nu_{dyn,K/\pi} = \frac{<N_{K}(N_{K}-1)>}{<N_{K}>^{2}}
+ \frac{<N_{\pi}(N_{\pi}-1)>}{<N_{\pi}>^{2}}
- 2\frac{<N_{K}N_{\pi}>}{<N_{K}><N_{\pi}>}\ ,
\label{nudyn}
\end{eqnarray}
A formula similar to (\ref{nudyn}) can be constructed for $p/\pi$ and $K/p$ ratio fluctuations. Additional information about $\nu_{dyn}$ can be found in \cite{nudyn2,starkpiprl,CPOD_proceeding}.
Earlier measurements of particle ratio fluctuations utilized the variable $\sigma_{dyn}$ \cite{NA49}. 
The two variables are related as $\sigma_{dyn}^{2} \approx \nu_{dyn}$. \cite{jeon,baym,sdasthesis}

\section{Experimental Analysis}

The data presented here for $p/\pi$, $K/p$, and $K/\pi$ dynamical fluctuations was acquired by the STAR experiment at RHIC from minimum bias (MB) Au+Au collisions at $\sqrt{s_{NN}}$ = 7.7, 11.5, 39, 62.4, and 200 GeV \cite{STAR}. Particle identification cuts can be found in \cite{starkpiprl,CPOD_proceeding}. 
Data was also analyzed using the recently completed Time of Flight (TOF) detector \cite{STARTOF}. 
TOF identified particles extended the $p_{T}$ range of pions and kaons from, $0.6 < p_{T} < 1.4$ GeV/$c$ and protons from $1.0 < p_{T} < 1.8$ GeV/$c$. We estimate that pion contamination of kaons is less than 3\% using a combination of the STAR TPC and TOF.

\section{Results and Discussion}

\begin{figure}
\subfigure[]{
\includegraphics[width=0.47\textwidth]{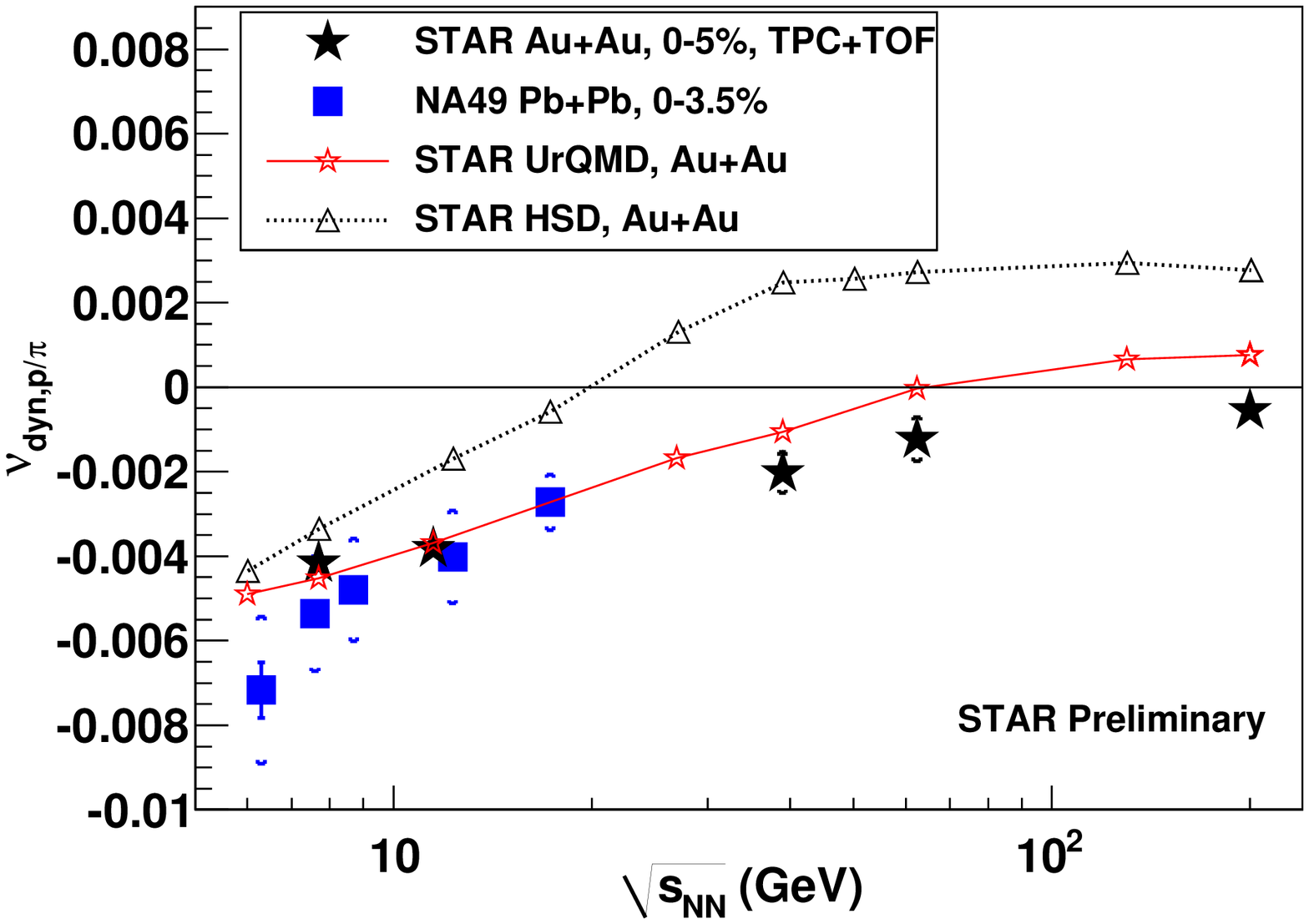} 
\label{ppi_excitation}
}
\subfigure[]{
\includegraphics[width=0.465\textwidth]{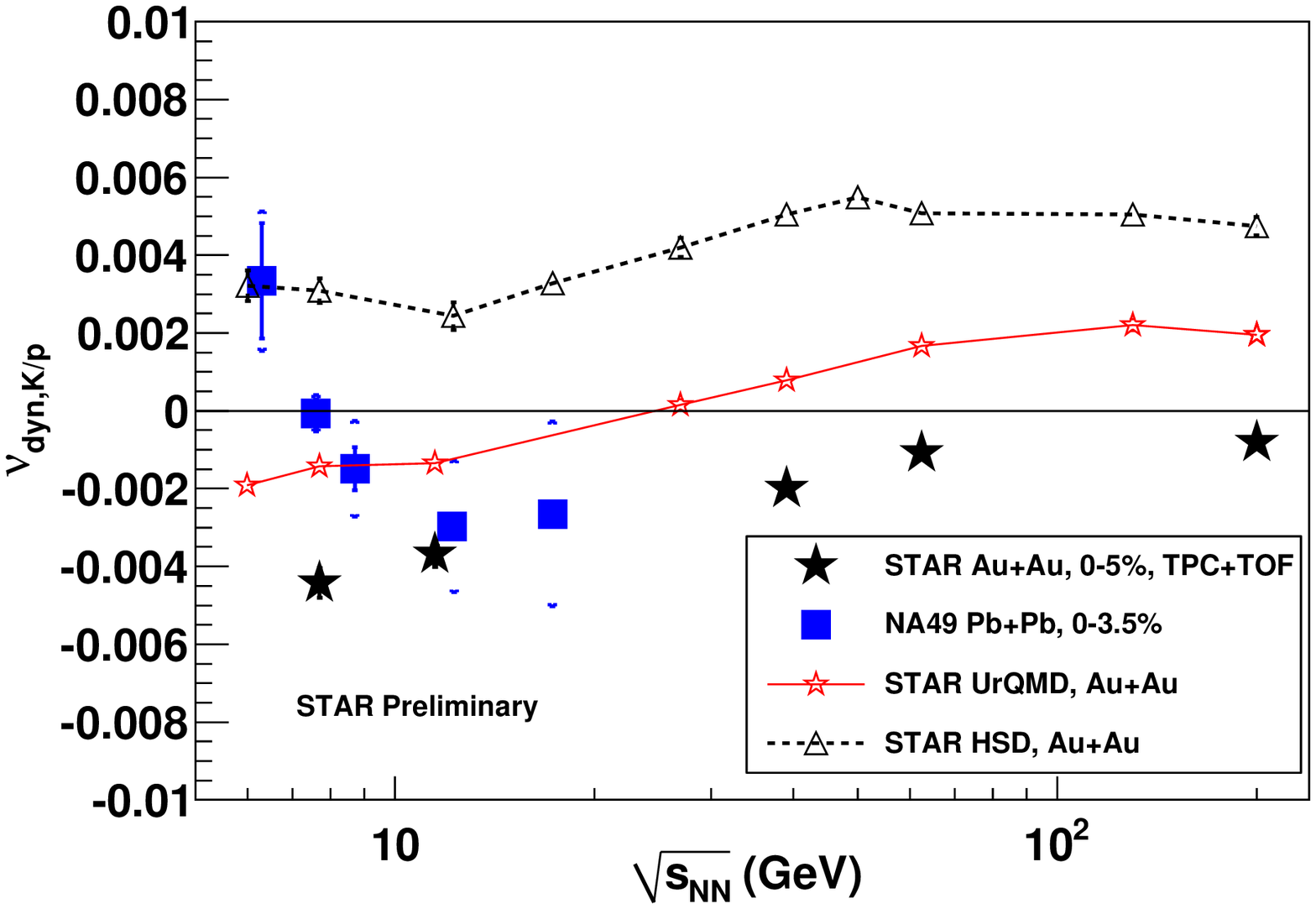} 
\label{kp_excitation}
}

\caption{Results for the measurement of $\nu_{dyn,p/\pi}$ (left) and $\nu_{dyn,K/p}$ (right) as measured by the STAR TPC+TOF (black stars) from central 0-5\% Au+Au collisions at $\sqrt{s_{NN}}$ = 7.7-200 GeV.  Also shown are results from the NA49 (solid blue squares) from central 0-3.5\% Pb+Pb collisions. Model predictions from UrQMD and HSD using the STAR experimental acceptance (red and black lines, respectively)  are also included.}
\label{asdf}
\end{figure}

\begin{figure}[]
\subfigure[]{
\includegraphics[width=0.51\textwidth]{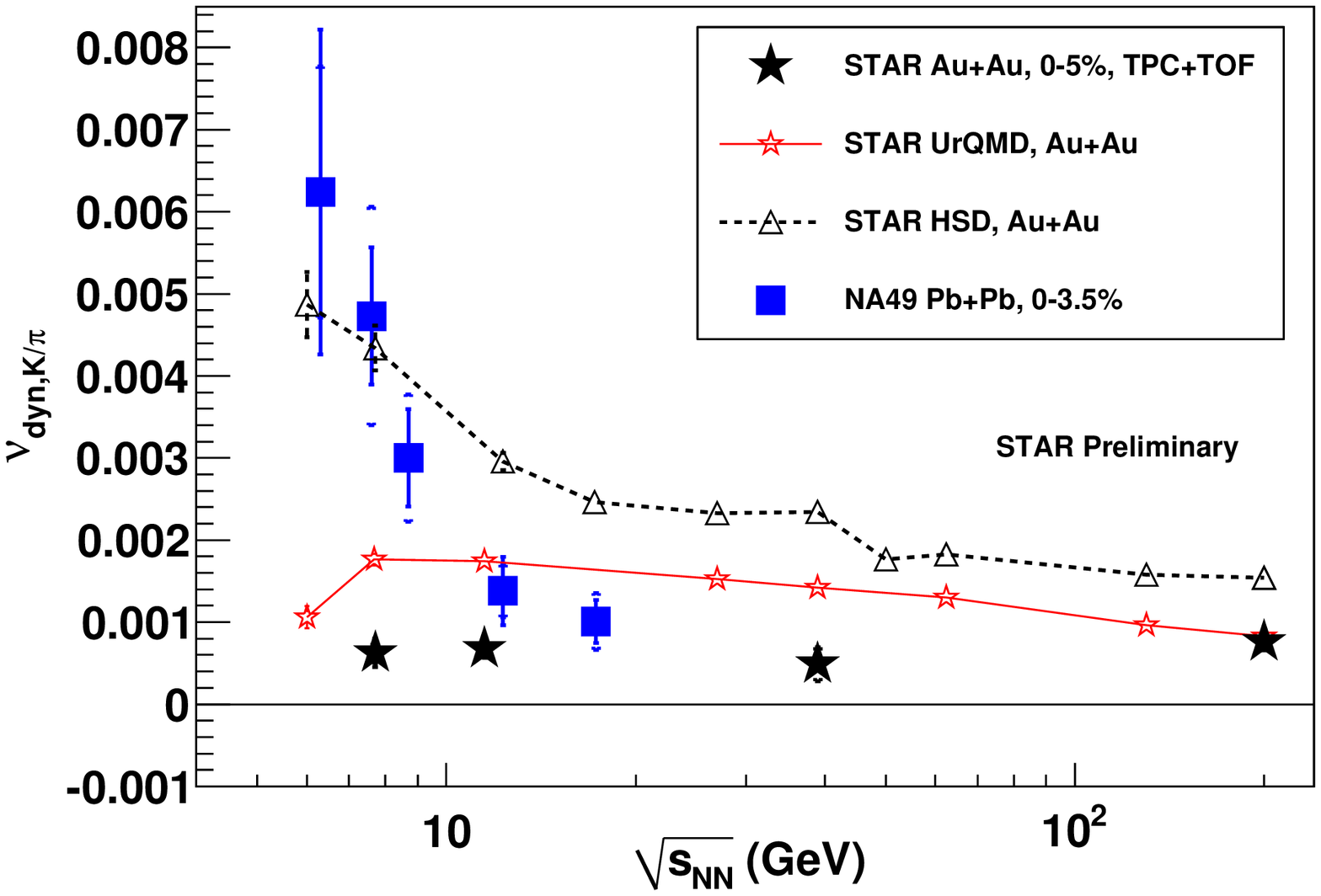} 
\label{kpi_excitation}
}
\subfigure[]{
\includegraphics[width=0.44\textwidth]{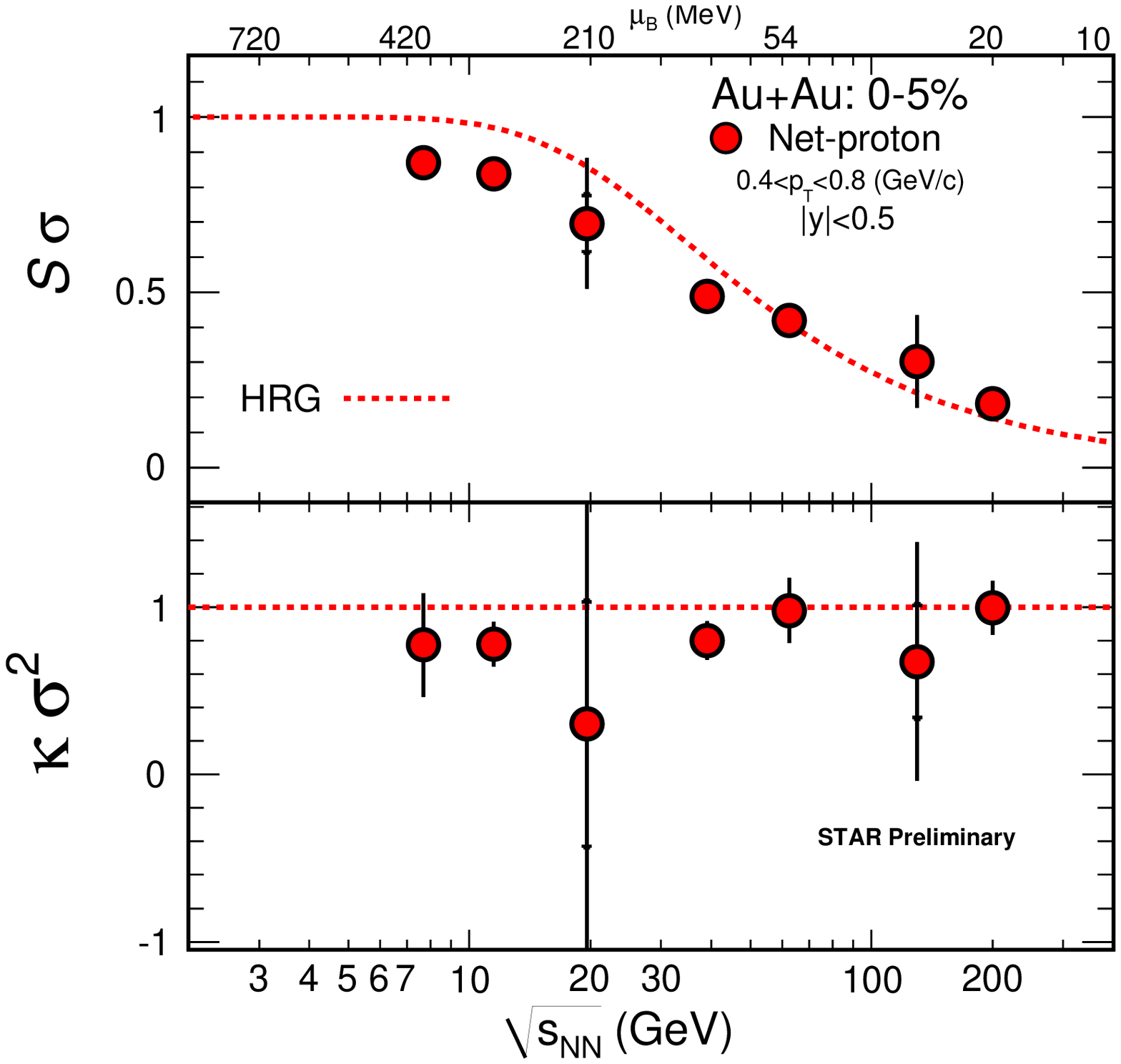} 
\label{netp}
}
\caption{Results for the measurement of $\nu_{dyn,K/\pi}$ (left) as measured by the STAR TPC+TOF (black stars) from central 0-5\% Au+Au collisions at $\sqrt{s_{NN}}$ = 7.7-200 GeV. 
Results for the measurement of net-proton Skewness*$\sigma$ and Kurtosis*$\sigma^{2}$ (right) as measured by the STAR experiment from central 0-5\% Au+Au collisions at $\sqrt{s_{NN}}$ = 7.7-200 GeV. The data is compared to the prediction of a Hadron Resonance Gas (HRG) model.}
\end{figure}

New results for dynamical particle ratio fluctuations from the RHIC energy scan are shown in Figures \ref{ppi_excitation}, \ref{kp_excitation}, and \ref{kpi_excitation}. Figure \ref{ppi_excitation} shows the measured dynamical $p/\pi$ fluctuations as a function of incident energy, expressed as $\nu_{dyn,p/\pi}$. The measured fluctuations from the STAR experiment are plotted as the black stars, from the NA49 experiment \cite{NA49_kpi_ppi} at the SPS as blue squares, and two transport model predictions from UrQMD and HSD (red and black lines, respectively), using the STAR experimental acceptance. All model predictions in this proceeding are processed through the same acceptance. The published NA49 results have been converted to $\nu_{dyn}$ using the relation $\sigma_{dyn}^{2} \approx \nu_{dyn}$. STAR measures a general trend from larger, negative values of $p/\pi$ fluctuations at lower energies, which then increases towards zero at $\sqrt{s_{NN}}$ = 200 GeV. These fluctuations are negative due to the dominance of the third term (correlated production, e.g. $\Delta^{++} \rightarrow p^{+} + \pi^{+}$) in $\nu_{dyn,p/\pi}$. There is good agreement between the two different experiments at the lowest energies measured by STAR ($\sqrt{s_{NN}}$ = 7.7 and 11.5 GeV). 

Figure \ref{kp_excitation} shows the measured dynamical $K/p$ fluctuations as a function of incident energy, expressed as $\nu_{dyn,K/p}$. Results from the NA49 experiment are from \cite{NA49_kp}. 
Dynamical $K/p$ fluctuations show an increase from larger, negative values toward zero as a function of increasing incident energy. This is qualitatively and quantitatively similar to the trend measured for dynamical $p/\pi$ fluctuations. 
The experimental trend observed by STAR at the lowest incident energies ($\sqrt{s_{NN}}$ = 7.7 and 11.5 GeV) is dramatically different than that measured by NA49. The STAR result is consistent with negative dynamical $K/p$ fluctuations down to $\sqrt{s_{NN}}$ = 7.7 GeV, while that from NA49 is consistent with zero and increases to a positive value at $\sqrt{s_{NN}}$ = 6.3 GeV. 

Figure \ref{kpi_excitation} shows the measured dynamical $K/\pi$ fluctuations as a function of incident energy, expressed as $\nu_{dyn,K/\pi}$. 
Results from the NA49 experiment are from \cite{NA49_kpi_ppi}. STAR measures dynamical $K/\pi$ fluctuations that are approximately independent of collision energy from $\sqrt{s_{NN}}$ = 7.7-200 GeV. This result is different than the observation of increasing fluctuations with decreasing incident energy observed by NA49. 
It is worth noting that the differences between the two experiments occur for dynamical ratio fluctuations involving kaons. There is good agreement between STAR and NA49 for $p/\pi$ dynamical fluctuations.

Figure \ref{netp} shows the experimental measurements for the products of the higher moments of the net-proton distribution. The higher moments of conserved quantities are predicted to be more sensitive to the correlation length of the system and may show a large change if the system passes through a critical point. The products of the Skewness*$\sigma$ ($S\sigma$) and Kurtosis*$\sigma^{2}$ ($\kappa\sigma^{2}$) are constructed to cancel volume effects \cite{Cheng}. For the net-proton distribution, it is observed that the products of the moments agree with a Hadron Resonance Gas (HRG) prediction \cite{Karsch_Redlich} above $\sqrt{s_{NN}}$ = 39 GeV and are below the prediction from $\sqrt{s_{NN}}$ = 7.7-39 GeV. These measurements are corrected for statistical effects related to the width of the centrality bin. 

%

\section{Summary}

New results from dynamical particle ratio fluctuations ($p/\pi$, $K/p$, and $K/\pi$) and the higher moments of net-proton distributions have been presented. 
Dynamical $p/\pi$ and $K/p$ fluctuations gradually increase from a larger negative value at $\sqrt{s_{NN}}$ = 7.7 GeV toward zero at $\sqrt{s_{NN}}$ = 200 GeV. Dynamical $K/\pi$ are positive and approximately energy independent from $\sqrt{s_{NN}}$ = 7.7-200 GeV. There are differences between the measured values of $K/p$ and $K/\pi$ fluctuations at $\sqrt{s_{NN}}$ = 7.7 GeV from the STAR and NA49 experiments. Results for the products of the higher moments of the net-proton distributions are in good agreement with the prediction of a Hadron Resonance Gas model at high energies, but deviations are seen at energies below $\sqrt{s_{NN}}$ = 62.4 GeV. Additional data at $\sqrt{s_{NN}}$ = 19.6 and 27 GeV will provide precision measurements of all fluctuation observables.


\section*{References}
\begin{thebibliography}{99}
\bibitem{Koch1}
  V.~Koch,
  arXiv:0810.2520 [nucl-th] (2008).
	
\bibitem{Stephanov}
M.~A.~Stephanov, 
Phys.\ Rev.\ Lett.\ {\bf 102}, 032301 (2009).
	
\bibitem{STARBES}
M.~M.~Aggarwal \emph{et al.} (STAR Collaboration),
arXiv:1007.2613v1 [nucl-ex]

\bibitem{kurtosis}
M.~M.~Aggarwal \emph{et al.} (STAR Collaboration),
Phys.\ Rev.\  Lett. {\bf 105}, 022302 (2010).

%
%
%
\bibitem{nudyn1}
C.~Pruneau, S.~Gavin and S.~Voloshin,
Phys.\ Rev.\  C {\bf 66}, 044904 (2002).

\bibitem{nudyn2}
J.~Adams \emph{et al.} (STAR Collaboration), 
Phys.\ Rev.\  C {\bf 68}, 044905 (2003).

\bibitem{starkpiprl}
B.~I.~Abelev \emph{et al.} (STAR Collaboration),
Phys.\ Rev.\  Lett. {\bf 103}, 092301 (2009).

\bibitem{CPOD_proceeding}
T.~Tarnowsky, 
arXiv:1101.3351 [nucl-ex] (2011).

\bibitem{NA49}
C.~Alt \emph{et al.} (NA49 Collaboration), 
Phys.\ Rev.\ C {\bf 79}, 044910 (2009).

\bibitem{jeon}
S.~Jeon and V.~Koch, 
arXiv:hep-ph/0304012v1 (2003).

\bibitem{baym}
G.~Baym, and H.~Heiselberg, 
Phys.\ Lett.\  B {\bf 469}, 7 (1999).

\bibitem{sdasthesis}
S.~Das, Ph.D. thesis, 2005.

\bibitem{STAR}
K.~H.~Ackermann \emph{et al.}, 
Nucl.\ Instr.\ and Meth.\  A {\bf 499}, 624 (2003).

\bibitem{STARTPC}
M.~Anderson \emph{et al.}, 
Nucl.\ Instr.\ and Meth.\  A {\bf 499}, 659 (2003).

\bibitem{STARTOF}
W.~J.~Llope (STAR TOF Group), 
Nucl.\ Instr.\ and Meth.\  B {\bf 241}, 306 (2005).

%
\bibitem{NA49_kpi_ppi}
C.~Alt \emph{et al.} (NA49 Collaboration),
Phys.\ Rev.\ C {\bf 79}, 044910 (2009).

\bibitem{NA49_kp}
T.~Anticic \emph{et al.} (NA49 Collaboration),
arXiv:1101.3250v2

\bibitem{Cheng}
M.~Cheng \emph{et al.}, 
Phys.\ Rev.\ D. {\bf 79}, 074505 (2009).

\bibitem{Karsch_Redlich}
F.~Karsch and K.~Redlich, 
Phys.\ Lett.\ B. {\bf 695}, 136 (2011).
	
\end{thebibliography}
\end{document}